\documentclass{emulateapj}
\usepackage{apjfonts}
\usepackage{epsfig}

\newcommand{\msun}{$\rm M_\odot$}
\newcommand{\dfn}{D$_{n}(4000)$}
\newcommand{\ha}{H$\alpha~$}
\newcommand{\hb}{H$\beta~$}

\slugcomment{To be published in Astrophysical Journal Letters}
\shorttitle{Dissecting Galaxy Colors}
\shortauthors{Johnson et al.}

\begin{document}

\title{Dissecting Galaxy Colors with \emph{GALEX}, SDSS, and \emph{Spitzer}}
\author{B. D. Johnson\altaffilmark{1}, D. Schiminovich\altaffilmark{1},   M. Seibert\altaffilmark{2}, M. A. Treyer\altaffilmark{2,3},S. Charlot\altaffilmark{4,5}, T. M. Heckman\altaffilmark{6}, D. C. Martin\altaffilmark{2}, S. Salim\altaffilmark{7}, G. Kauffmann\altaffilmark{4}, L. Bianchi\altaffilmark{8}, J. Donas\altaffilmark{5}, P. G. Friedman\altaffilmark{2}, Y.-W. Lee\altaffilmark{9}, B. F. Madore\altaffilmark{10}, B. Milliard\altaffilmark{5}, P. Morrissey\altaffilmark{2}, S. G. Neff\altaffilmark{11}, R. M. Rich\altaffilmark{7}, A. S. Szalay\altaffilmark{2}, K. Forster\altaffilmark{2},  T. A. Barlow\altaffilmark{2}, T. Conrow\altaffilmark{2}, T. Small\altaffilmark{2}, and T. K. Wyder\altaffilmark{2}}

\affil{}

\altaffiltext{1}{Department of Astronomy, Columbia University, 550 West 120th Street, New York, NY 10027; bjohnson@astro.columbia.edu}
\altaffiltext{2}{California Institute of Technology, MC 405-47, 1200 East California Boulevard, Pasadena, CA 91125}
\altaffiltext{3}{Laboratoire d'Astrophysique de Marseille, BP8, Traverse du Siphon, F-13376 Marseille, France}
\altaffiltext{4}{Max-Planck Institute fur Astrophysik, D-85748, Garching, Germany}
\altaffiltext{5}{Institut d'Astrophysique de Paris, CNRS, 98 bis Boulevard Arago, F-75014 Paris, France}
\altaffiltext{6}{Department of Physics and Astronomy, The Johns Hopkins University, Homewood Campus, Baltimore, MD 21218}
\altaffiltext{7}{Department of Physics and Astronomy, University of California, Los Angeles, CA 90095}
\altaffiltext{8}{Center for Astrophysical Sciences, The Johns Hopkins` University, 3400 N. Charles St., Baltimore, MD 21218}
\altaffiltext{9}{Center for Space Astrophysics, Yonsei University, Seoul 120-749, Korea}
\altaffiltext{10}{Observatories of the Carnegie Institution of Washington, 813 Santa Barbara St., Pasadena, CA 91101}
\altaffiltext{11}{Laboratory for Astronomy and Solar Physics, NASA Goddard Space Flight Center, Greenbelt, MD 20771}

\begin{abstract}

We combine data from SDSS and the \emph{GALEX} and \emph{Spitzer} observatories to create a sample of galaxies observed homogeneously from the UV to the Far-IR.  This sample, consisting of $\sim$460 galaxies observed spectroscopically by SDSS provides a multiwavelength (0.15-24 $\mu$m) view of obscured and unobscured star formation in nearby ($z<0.3$) galaxies with SFRs ranging from 0.01 to 100 M$_\odot$ yr$^{-1}$.  We calculate a robust dust measure from the infrared to UV ratio (IRX) and explore the influence of star formation history (SFH) on the dust-UV color relation (i.e. the IRX-$\beta$ relation). We find that the UV colors of galaxies are only weakly dependent on their SFH as measured by the 4000\AA~ break. However, we find that the contributions of dust and SFH are distinguishable when colors at widely separated wavelengths (e.g. 0.23-3.6\micron) are introduced. We show this explicitly by recasting the IRX-$\beta$ relation as a more general IRX-SFH-color relation, which we examine in different projections.  We also determine simple fits to this relation.

\end{abstract}

\keywords{galaxies:evolution --- dust:extinction --- galaxies:ISM --- ultraviolet:galaxies --- infrared:galaxies} 

\section{Introduction}
\label{sec:intro}

\setcounter{footnote}{0}

The color of a galaxy is determined primarily by its star formation history (SFH) and the amount of dust attenuation present, with significant additional contributions from metallicity and dust geometry.  An empirical relation between SFH, dust attenuation, and color is therefore a useful constraint on models of galaxy formation, and can aid in the interpretation of high redshift galaxy observations where measurements are difficult.

Here we investigate such an empirical relation, using a representative sample of galaxies observed from the ultraviolet to the infrared by \emph{GALEX}, SDSS, and \emph{Spitzer}. The long wavelength coverage allows us to construct  simple but robust measures of dust attenuation that are relatively free of a dependence on SFH.  Similarly, the SDSS spectroscopy allows us to measure SFH diagnostics -- we use a 4000\AA~ break measure -- that are largely free of a dependence on dust attenuation.

Earlier studies of the effect of dust on galaxy colors have focused on the ultraviolet (UV, $\lambda\lambda\sim 1400-2500\AA$) colors of galaxies.  A primary reason for this is that the intrinsic, underlying UV color (before attenuation) is relatively insensitive to the SFH of the galaxy, when compared to the effects of dust attenuation.  Also, studies of the highest redshift galaxies are often restricted to the restframe UV so that estimates of dust attenuation must be made using UV colors.  \citet{calzetti94} derived the effective attenuation properties of dust in star-bursting galaxies by comparing the total dust absorption -- measured from the ratio of dust emission to UV emission -- to the change in UV color of a sample of starforming galaxies (the so-called IRX-$\beta$ relation). This locally derived relation has been used extensively to determine the dust attenuation in galaxies at higher redshift using their UV colors (e.g. \citet{MHC}). However, more recent studies \citep{bell02a,seibert05a, buat05} have shown that there is significant scatter in this relation, especially when less rapidly starforming galaxies are included.  \citet{kong04} show, using stellar population synthesis models with significant recent bursts, that both SFH and dust are expected to affect the UV color of normal galaxies since at a constant value of dust attenuation galaxies with an older stellar population should appear redder in the UV because of their redder intrinsic UV spectra.  They thus propose that a measure of SFH -- specifically the 4000\AA~ break -- can be used to explain the scatter in the IRX-$\beta$ relation. 

\begin{figure*}
\epsscale{1.05}
\plotone{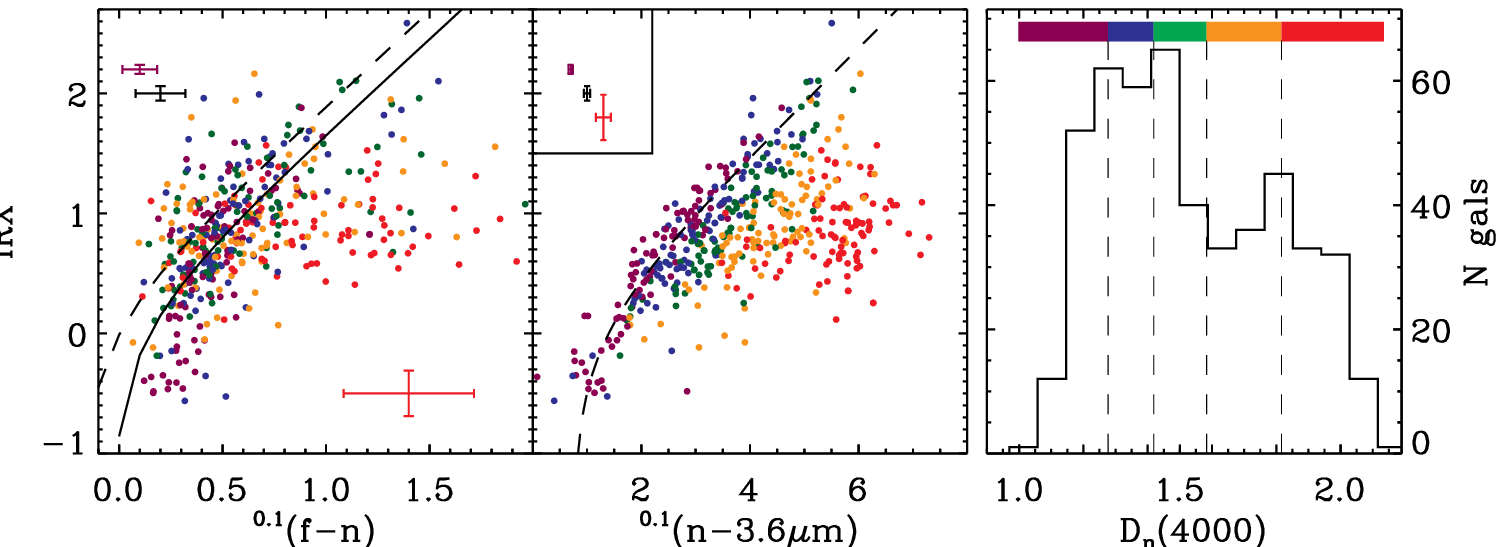}
\figcaption{ IRX vs. color of the sample for different bins of \dfn.  {\it Left}: $f-n$ color. Lines show the relations of \citet{MHC} (dashed) and \citet{seibert05a} (solid) where the conversions from A$_{fuv}$ and $\beta$ are from \citet{seibert05a}. {\it Middle}: The long-baseline $n-3.6\micron$ color. Dashed line gives the fit from the 3rd row of Table~\ref{table:fit} for \dfn~$=1.25$. {\it Right:} The distribution of \dfn for the sample, with dashed lines and color bar showing the quintiles used to color code the sample galaxies.  Error bars show the median measurement errors (excluding systematic effects) for the entire sample (\emph{black}) and the lowest and highest quintiles of \dfn~(\emph{purple} and \emph{red} respectively).  Note the clear separation of galaxies with different \dfn~ in the middle panel, while for the $f-n$ color (left) such a separation is marginal.  
\label{fig:bigfig}}
\end{figure*}

\section{Data}
\label{sec:data}
We have constructed a sample of galaxies observed from the UV to the mid/far-infrared.  Our primary data set is the sample of local ($z<0.3$) galaxies observed spectroscopically by SDSS and analyzed by \citet{kauffmann03a} and \citet{brinchmann04} (hereafter SDSS/MPA galaxies). The UV data is taken from pipeline processed \emph{GALEX} observations of the Lockman Hole, with exposure times of $\sim 1.5$ ks.  The pipeline-produced \emph{GALEX} catalogs are searched for objects within 3" of the SDSS/MPA galaxy locations, and the nearest object is taken as the match to the SDSS galaxy. UV flux measurements are made in elliptical Kron apertures. In the optical we use SDSS petrosian magnitudes. The infrared data is provided by the SWIRE \emph{Spitzer} observations of the Lockman Hole \citep{swire}.  We have performed aperture photometry in the SWIRE team processed\footnote{http://swire.ipac.caltech.edu/swire/astronomers.html for a description of the SWIRE image processing} 3.6 through 7.8\micron~ IRAC images and 24\micron~ MIPS images at the location of each of the SDSS/MPA galaxies, using a 7" radius aperture (12" at 24 \micron).  The fluxes are then aperture corrected to total magnitudes.  Systematic errors in IR flux due to calibration uncertainty, aperture corrections, and the resolved nature of many of the sources amount to $\sim 30\%$. 

The resulting UV through 3.6\micron~ magnitudes are $K$-corrected to $z=0.1$ (e.g. $^{0.1}u$, $^{0.1}g$, etc.) using the method of \citet{blanton_k}. At longer wavelengths dust emission becomes more important than stellar emission, and we use a different method to `$K$-correct' the data:  we choose the best fitting redshifted \citet{dale01} model IR SED, on the basis of the observed 8 to 24\micron~ flux ratio.  This SED is then normalized using the measured 24\micron~ flux, and the integrated far-infrared (8-1000\micron) dust luminosities ($L_{dust}$) are derived.  Note that the different \citet{dale01} SEDs have $L_{24\micron}/L_{dust}$ ratios that are different by a factor of up to five. We have checked that our results would not change significantly if we use the model SEDs of \citet{devriendt} (see \citet{papovich02} for a detailed discussion of predicting IR luminosities from \emph{Spitzer} data).  

Our final selection of galaxies consists of those SDSS/MPA galaxies with detections in the Far-UV ($f$, $\lambda=1516$\AA) and Near-UV ($n$, $\lambda=2267$\AA) through 24\micron~ bands. This is a total of 467 of the 645 SDSS/MPA galaxies within the $\sim9$ deg$^{2}$ of SDSS/\emph{GALEX}/\emph{Spitzer} overlap.  These galaxies have a stellar mass range of $\sim 10^{7}-10^{11.5}$ \msun~ and a SFR range of $\sim 0.01-100$ \msun$\mbox{yr}^{-1}$, as determined by \citet{kauffmann03a} and \citet{brinchmann04} from the optical spectra and photometry.  To simplify the present analysis we do not consider upper limits or selection effects except to note that we are biased against galaxies with very low $L_{fuv}$ or $L_{dust}$.  Nevertheless, we recover a significant number of galaxies which appear `old' and elliptical but have low levels of UV and IR emission.

\section{Analysis}

\subsection{Dust Indicators}

One of the primary motivations for compiling the sample of galaxies described above was to construct a robust and model independent measure of dust attenuation, the so-called infrared excess (IRX, see \citet{gordon00} for a discussion of the relation of infrared excess to UV attenuation).  We adopt the definition $\mbox{IRX}=\log (L_{dust}/ L_{fuv})$ where $L_{dust}$ is the 8-1000\micron~ dust luminosity as determined above and $L_{fuv}=\nu L_{\nu}$ is the luminosity in the  $^{0.1}f$ band ($\nu\approx c/1400$\AA). A second available dust indicator is the \ha to \hb decrement (see \citet{kennicutt98} and references therein), measured from the SDSS spectra.  This dust measure is only well defined for those galaxies with strong emission lines, and cannot be easily compared to global galaxy measures due to spectroscopic aperture effects.  We do not consider it in the present study.


\subsection{SFH Indicators}
\label{sec:sfh}
We take the strength of the 4000\AA~ break in galaxies as an indicator of SFH \citep{balogh99,brinchmann04,macarthur05}.  This has been measured from the deredshifted SDSS spectra by \citet{kauffmann03a} using the ratio of the flux in two narrow bands ($\Delta\lambda = 100\AA$) centered at 4050\AA~ and 3900\AA.  This narrowband color, \dfn, is less sensitive to reddening by dust than broadband colors. It is not, however, completely insensitive to dust effects \citep{macarthur05}. Note that \dfn~ is only measured within the 3" SDSS aperture, which can cause an overestimate of the integrated \dfn~ for galaxies with moderate bulge to disk ratios.  We do not consider in detail the relation between \dfn~ and more physical measures of galaxy age or SFH (e.g. the ratio of current to past averaged star formation rate or the specific star formation rate), which is metallicity dependent and requires population synthesis modeling. 

\section{Results: The Dust-SFH-Color Relation}
\subsection{IRX vs. Color Binned by \dfn}
\label{sec:irx_color}

Models of galaxy spectra have suggested that the UV color of galaxies can be decomposed into contributions from dust and SFH, though the effect of dust is the dominant contribution \citep{kong04}. In Figure \ref{fig:bigfig} we show the relation between IRX and color for different ranges of \dfn.  We find a marginal dependence of the IRX vs. $f-n$ relation (analogous to the IRX-$\beta$ relation) on \dfn. Only the very oldest galaxies (red points) have systematically redder $f-n$ for the same value of IRX, but also have larger errors.  For galaxies with \dfn$\lesssim 1.8$ there appears to be little dependence of the scatter on galaxy SFH.  This is similar to the result of \citet{seibert05a}, who use the $n-$K color as a measure of SFH (but see below for the effect of dust on the $n-$NIR color). 

This result can be reconciled with the models of \citet{kong04} in several ways.  First, some stars contributing to the dust heating (and therefore the IR emission) may not be contributing to the UV emission or, conversely, some stars contributing to the UV emission are not contributing to the IR emission (i.e. a decoupling of IR and UV emission). This may occur for various reasons including star-dust geometry (e.g. \citet{calzetti05,thilker05}) and contributions to dust heating from older stellar populations. Related to this, \dfn~ may not be measuring the relevant timescale for changes in the UV spectrum.   The SFH diagnostic that is likely to be most relevant to changes in the UV spectral shape is the ratio of current SFR to the SFR averaged over the last 100 Myr \citep{calzetti05}, which is not probed well by \dfn. Second, weak AGN may affect the UV colors. Third, the entire range of $f-n$ color is $\lesssim 2$ mag, compared to a median error of $\sim0.25$ mag.  The scatter due to measurement error may obscure a trend in the scatter with SFH. The results of \citet{kong04} were based on a library of 95000 model spectra, which may not be well represented by our sample.  These and other possibilities will be investigated in future work. 

For $n-3.6\micron$ color it is easy to see the effect of SFH that was predicted by \citet{kong04} for the $f-n$ color.  At a given low \dfn~ (i.e. younger mean stellar age, purple points) the relation between dust and color is clear, and has low scatter.  This relation is closely related to the effective attenuation curve of the dust in these galaxies \citep{calzetti94, MHC}. For intermediate \dfn~ (i.e.\ intermediate mean stellar age, orange or green points) the relation between dust and color persists with low scatter, but the entire relation is shifted to redder color.  This is presumably because of the redder intrinsic spectrum of an older stellar population, on top of which the effect of dust attenuation on color persists relatively unchanged. For $n-3.6\micron~$ the ratio of the median error to the range in color is significantly smaller than for $f-n$, making trends with color easier to see.  

At the largest \dfn~ the scatter in the IRX-color relation increases -- this may be due to several causes.  First, there are larger errors in IRX (and color) for these galaxies, which have relatively little star formation and are systematically dimmer in $f$ and at 24\micron. Second, the effects of metallicity on \dfn~ become more pronounced at high \dfn.  Third, some UV emission may be due to evolved populations (e.g. BHB stars).  Finally, these galaxies may well host AGN that affect the IR and/or UV emission, changing IRX.

\subsection{\dfn~ vs. Color Binned by IRX}
\begin{figure}
\plotone{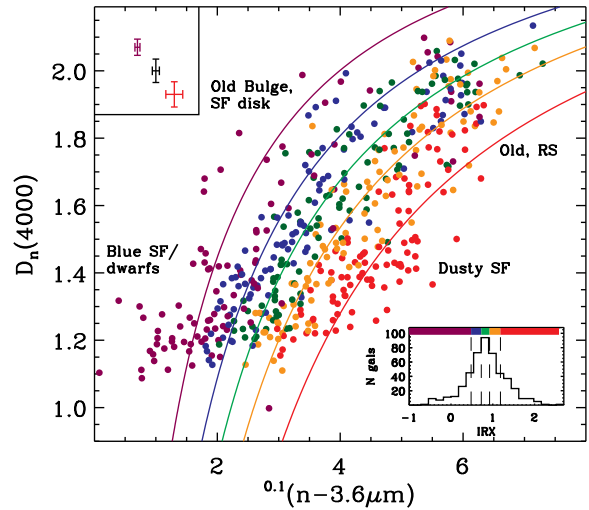}
\epsscale{1.1}
\figcaption{\dfn~ vs. $n-3.6\micron~$ color for the sample, with color coding by IRX. The colored lines show a fit to the relation (3rd row from Table \ref{table:fit}), where each line represents the fitted relation for the median value of IRX in the bin of the corresponding color. Labels indicated the type of galaxies typically occupying different regions of the diagram (SF-star formation; RS- red sequence). Error bars show the median measurement errors (excluding systematic effects) for the entire sample (\emph{black}) and the lowest and highest bins of IRX~(\emph{purple} and \emph{red} respectively).  {\it Inset, Lower Right}: Distribution of IRX for the sample.  Dashed lines and color bars show the bins used to color code the the sample.
\label{fig:model}}
\end{figure}

In Figure \ref{fig:model} we present another projection of the dust-SFH-color relation.  Here we show \dfn~ as a function of $n-3.6\micron$ color for different ranges of IRX. IRX appears to be well determined using just $n-3.6\micron$ and \dfn.  The relation between \dfn~ and $n-3.6\micron~$ shifts to redder color for galaxies with more dust, while the slope of the relation remains nearly constant. Galaxies in the lower right of this plot are predominantly dusty star-forming galaxies.  Galaxies in the lower left are typically blue star-forming galaxies or dwarfs/irregulars.  Galaxies in the upper right of Figure \ref{fig:model} are red-sequence galaxies composed primarily of old stars, with very low levels of UV and IR emission.   



\subsection{Fits to the Relation}

\begin{figure}
\plotone{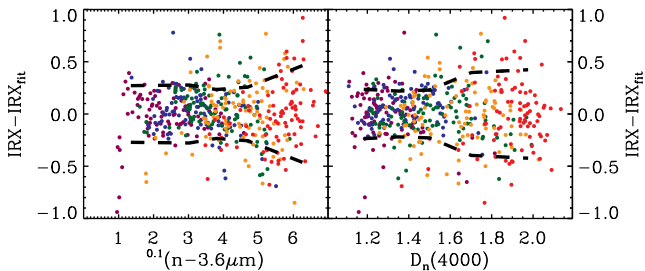}
\figcaption{Residuals in IRX from a simple functional fit of the dust-SFH-color relation (3rd row of Table \ref{table:fit}). Dashed lines show the smoothed rms dispersion of the residuals. {\it Left}: Plotted against the $n-3.6\micron$ color. The color of each point indicates its \dfn~ quintile. {\it Right}: Plotted against \dfn. The color of each point indicates its $n-3.6\micron$ color quintile.
\label{fig:fit}}
\end{figure}

We have conducted quantitative parametric fits to the relation between dust attenuation, SFH, and the $n-3.6$\micron~ color. Considering a simple effective attenuation law we are motivated to consider polynomial fits to the color ($c$), \dfn, and an expression of IRX that is linearly related to color $\widehat{A}_{IRX}=2.5\log(10^{IRX-\log BC_{*}}+1)$ where $BC_{*}=1.68$ is the bolometric correction to the UV and the bolometric correction to the IR has been made in \S\ref{sec:data} (see \citet{MHC,gordon00,seibert05a} for a discussion of the relation between IRX and the true UV attenuation $A_{FUV}$). We treat the color and \dfn~ as independent variables (since the errors are much smaller than for IRX).  We assume either a linear or quadratic form for the relation between color and \dfn. The results are given in Table \ref{table:fit}.  Examination of the residuals shows that a cross term ($($\dfn$)*c$) is required, and fits including such a term are also given in Table~\ref{table:fit}.  The fit from the 3rd row of Table~\ref{table:fit} is overplotted in Figure~\ref{fig:model}, and the residuals in IRX as a function of color and \dfn~ are shown in Figure~\ref{fig:fit}. Figure \ref{fig:fit} shows that the fit is poorer for redder colors and larger \dfn.  This may be due to the effects listed in \S\ref{sec:irx_color} for the $n-3.6$\micron~ color, as well as the effect of aperture on \dfn~ (\S\ref{sec:sfh}).  At bluer colors ($n-3.6$\micron $<4$) and smaller \dfn~ the residuals are lower, $\sim0.4$ rms in $\widehat{A}_{IRX}$. Table \ref{table:fit} also includes fits using the $n-r$ color since this is more easily measured for a large sample of GALEX observed SDSS galaxies (e.g. \citet{salim}).  The behavior is similar to the $n-3.6$\micron~ color though the residuals are larger.


\begin{deluxetable}{cccccccc}
\tablecolumns{7} 
\tablewidth{0pc} 
\tablecaption{Fits to the Dust-SFH-color Relation of the form\tablenotemark{a} $\widehat{A}_{IRX}=\mbox{A}+\mbox{B}x+\mbox{C}x^{2}+\mbox{D}y+\mbox{E}xy$ where $x=($\dfn$-1.25)$ \label{table:fit}} 
\tablehead{ 
\colhead{A} & \colhead{B} & \colhead{C} & \colhead{D} & \colhead{E} & \colhead{RMS ($\widehat{A}_{IRX}$)} & \colhead{RMS (IRX)}}
\startdata 
\multicolumn{7}{c}{$y=~^{0.1}(n-3.6$\micron$)-2$} \\
\hline
 1.31 & -3.46  &  --   & 0.84 &  --   & 0.57 & 0.33\\
 1.21 & -1.59  & -2.73 & 0.83 &  --   & 0.55 & 0.31\\
 1.08 & -1.80  &  --   & 1.03 & -0.72 & 0.51 & 0.31\\
 1.07 & -2.39  & 1.59  & 1.10 & -0.94 & 0.50 & 0.32\\
\hline
\multicolumn{7}{c}{$y=~^{0.1}(n-r)-2$} \\
\hline
 1.60 & -3.27 &  --   & 0.87 &  --   & 0.75 & 0.43\\
 1.43 & -0.58 & -4.12 & 0.89 &  --   & 0.70 & 0.39\\
 1.27 & -1.56 &  --   & 1.35 & -1.24 & 0.65 & 0.38\\
 1.25 & -2.91 & 3.30  & 1.56 & -1.82 & 0.64 & 0.39
\enddata
\tablenotetext{a}{$\widehat{A}_{IRX}=2.5\log(\frac{1.68L_{fuv}+L_{dust}}{1.68L_{fuv}})$ is only an approximation of the FUV attenuation $A_{FUV}$ \citep{MHC,gordon00}.}

\end{deluxetable}



We have used a simple treatment of attenuation to motivate our fits that does not self consistently consider, e.g., the heating of dust by old stars. Also, \dfn~ is subject to aperture effects that may affect the fits. It is thus difficult to interpret the fit coefficients as physical parameters. The fits in Table \ref{table:fit} are referenced to \dfn$=1.25$, $c=2.0$ because this region is typical of `blue sequence' galaxies.  The constant A term thus gives the typical $\widehat{A}_{IRX}$ for such galaxies. The B term (with contributions from the C and E terms when present) gives the trend in $\widehat{A}_{IRX}$ with \dfn, and is negative since galaxies that are intrinsically redder have less attenuation for a given color. Similarly, D gives the trend in $\widehat{A}_{IRX}$ with color, and is positive as a redder galaxy at a given \dfn~ has more attenuation.  


\section{Conclusions}

While we recover only a weak SFH dependence for the short wavelength-baseline UV color, we have shown that long wavelength-baseline colors --  specifically $n-3.6\micron$ -- can be decomposed into contributions from dust and SFH with low scatter.  This is the dust-SFH-color relation.  Such a decomposition is possible due to the use of a relatively dust-insensitive (though perhaps metallicity dependent) SFH indicator in combination with a robust measure of the dust attenuation in galaxies. At large \dfn~ the relation between dust, SFH, and color is more scattered, suggesting that an additional parameter may be necessary to explain the color of such galaxies, or that our SFH and/or dust indicators become less reliable here.  Deep \emph{Spitzer} IRAC data, when combined with optical data, is ideally suited to measuring the $n-3.6\micron$ color for galaxies at $z=0-2$, allowing similar analyses for galaxies at much earlier epochs (e.g. \citet{kriek06} and \citet{reddy_dust}). 

\acknowledgements
The MPA/JHU collaboration for SDSS studies has very generously made their catalogs publicly available.  The publicly available \emph{Spitzer} data obtained and reduced by the SWIRE team have been essential to this work. We gratefully acknowledge NASA's support for construction, operation, and science analysis for the \emph{GALEX} mission, developed in cooperation with the Centre National d'Etudes Spatiale of France and the Korean Ministry of Science and Technology. BDJ was supported by NASA GSRP Grant NNG05GO43H. 


\clearpage

\clearpage




\end{document}